\documentclass[aps,prd,superscriptaddress,showpacs,twocolumn,floatfix]{revtex4}
\usepackage{graphicx}
\newcommand{\ba}{\begin{eqnarray}}
\newcommand{\ea}{\end{eqnarray}}
\newcommand{\beqs}{\begin{eqnarray}}
\newcommand{\eeqs}{\end{eqnarray}}

\begin{document}

\title{GPDs of hadrons and elastic pion-nucleon scattering}


\author{  O.V. Selyugin\footnote{selugin@theor.jinr.ru} } 
\address{\it Bogoliubov Laboratory of Theoretical Physics, \\
Joint Institute for Nuclear Research,
141980 Dubna, Moscow region, Russia }



\begin{abstract}
  The pion  structure  is represented by generalized parton distribution functions (GPDs).
   The momentum transfer dependence of GPDs of the pion   was obtained
   on the basis of the form of GPDs of the nucleon  
   in the framework of the high energy generalized structure  (HEGS) model.
   To this end,  different forms of  PDFs of the pion of various Collaborations  were examined
    with taking into account the available experimental data on the pion form factors.
    As a result,
      the electromagnetic and gravitomagnetic form factors of the pion were calculated.
  They were used in the framework of the HEGS model
  with the electromagnetic and gravitomagnetic form factors of the proton
   for  describing
    pion-nucleon   elastic scattering in a wide energy and momentum transfer region
  with a minimum of fitting parameters. The properties of the obtained scattering amplitude were
  analyzed.
\end{abstract}

\maketitle %

\section{Introduction}

          The study of the particle structure  is one of the old and long-standing
                   problems in modern physics.
  The main step was made by introducing the parton picture of  hadrons.
   Now many collaborations have
  obtained some forms of the parton distribution functions (PDF) using the recent data obtained at HERA and LHC
  in  deep inelastic scattering. Besides this main point of the modern picture of the hadron structure, which depends
  only on the Bjorken longitudinal variable $x$, there were introduced a number of  other more complicated
  structure functions, for example,  the generalized parton distribution functions (GPDs)
  (which  depend on $\ x$,  momentum transfer $t$
  and the skewness parameter $\xi$),
   transverse momentum distributions (TMDs)  functions  (which  depend on $x$, inner
  momentum transfer $k$ and skewness parameters $\xi$) and many others.
    Now we have  more generalized parton distributions which depend on  different variables 
     $GTMDs(x,\vec{k},\xi,\vec{\Delta)}$,  generalized transverse momentum dependent
     distributions of    partons  \cite{Meis-08,Lorce-13,Burk-15}.
       They are  parameterized by the unintegrated off-diagonal     quark-quark correlator
     depending on the three-momentum $\vec{k}$ of the quark and on the four-momentum, 
     which is transferred by the probe to the hadron. Taking $\Delta=0$,  we can obtain 
     $TMD(x,\vec{k})$ the transverse momentum-dependent parton distribution. In another way,
     after integration over $\vec{k}$ we obtain $GPDs(x,\xi,\Delta)$, generalized parton    distributions.

     The remarkable property of $GPDs$ is that the integration of  different momenta of
     GPDs over $x$ gives us  different hadron form factors  \cite{Mil94,Ji97,R97}.
     The $x$ dependence of $GPDs$ is   determined, in most part,  by the standard
    PDFs, which are obtained by many Collaborations from the analysis of
     deep-inelastic processes.
    Specific reactions   can be related with  different form factors.
  For example,  strong hadron-hadron scattering can be proportional to
   the gravitomagnetic form factor or the matter distribution of  hadrons, and
       the Compton scattering is described by the Compton form factors.
     Hence, the generalized parton distributions reveal themselves as a bridge between the data 
  on the inelastic reaction 
  and  the recent data on the elastic hadron-hadron cross section.
 Many different forms
  of the   
  momentum transfer dependence of GPDs were proposed.
   In the quark diquark model \cite{Liuti1} the form of  GPDs
   consists of three parts - PDFs, function distribution and the Regge-like function.
 In other works (see e.g. \cite{Kroll04}),
  the description of the $t$-dependence of  GPDs  was developed
  in a  more complicated picture using the polynomial forms with respect to $x$.


    Note that  functions like GPDs(x,t, $\xi=0$) were already used in the old
    "Valon" model proposed by Sanielevici and Valin in 1986  \cite{San-Val}.
  In that model, the hadron elastic form factor was obtained by the integration function
  $L(x) G(x,t)$ where $L(x)$ corresponds to the parton function and $G(x,t)$ corresponds  to an
  additional function which  depends on the momentum transfer and $x$. In modern   language,
  this exactly corresponds to  GPDs.
 The recent results from the LHC  gave  plenty of new information
  about the elastic and deep-inelastic  processes, which raised  new questions in the study of the structure of hadrons.

          In the paper, we analyze the hadrons structure, which is presented by the GPDs of hadrons.
          In Sec. II, the momentum transfer dependence of GPDs of hadrons obtained in the framework of the high energy generalized structure (HEGS) model is discussed.
       It is very important to check the obtained $t$-dependence of GPDs,
 as it  determined the $t$-dependence of the gravitomagnetic form factor of nucleons, which in turn impact on momentum transfer dependence of the differential cross sections.
   As an example, in Sec. III we calculated by integration
    other Mellin moments of GPDs which give us the corresponding Compton form factors and transition magnetic form factor. Comparing  the corresponding cross sections determined by Compton form factors and transition magnetic form factor with the existing experimental data gives us additional support of the obtained $t$-dependence of GPDs.
          In Se. IV, a short review of the  results of the HEGS model for the nucleon structure and nucleon-nucleon scattering is presented.
          In Sec. V, the GPDs of the pion
          are determined, and on their basis the electromagnetic and gravitomagnetic pion form factors are calculated.
          In Sec. VI, the obtained nucleon and
          pion form factors are used in the framework of the HEGS model for pion-nucleon elastic scattering. The conclusion is presented in the final section.

 \section{Momentum transfer dependence of GPDs of nucleon}

 In    \cite{ST-PRDGPD}, the standard Gaussian ansatz of the
   $t$-dependence of  GPDs is chosen  in a simple form
\ba
 {\cal{H}}^{q} (x,t) \  = q(x) \   exp [  a_{+}  \
   f(x) \ t ],                                    
   \label{GPD0}
\ea
  with $f(x)= (1-x)^{2}/x^{\beta}.$
The isotopic invariance can be used to relate the proton and neutron GPDs.
   The complex analysis of the corresponding description of the electromagnetic form factors of the proton and neutron
    by  different  PDFs sets  (24 cases) was carried out in \cite{GPD-PRD14}. These
   PDFs include the  leading order (LO), next leading order (NLO) and next-next leading order (NNLO)
   determination of the parton distribution functions.
   They used  different forms of the $x$ dependence of  PDFs. 
     A slightly complicated  form of GPDs was taken into acount
     in comparison with the equation used in     \cite{ST-PRDGPD},  
   but it is the simplest one as compared to other works (for example, \cite{DK-13},
 where   $f(x,t)_{q}$ was chosen in the form with different $x$ dependence,
 six parameters control
  the small $x$ behavior of these functions, whereas their behavior at large $x$ is controlled
   other six parameters.
   Note, that in \cite{Yuan-04}, it was shown that at large $x \rightarrow 1$ and momentum transfer the behavior of GPDs
requires a larger power of $(1-x)^b$ in the $t$-dependent exponent.
\ba
{\cal{H}}_{u} (x,t) \  &=& q_{u}(x)  \   e^{2 a_{H}   f_{u}(x)  \ t };  \ \ \ \\ \nonumber
{\cal{H}}_{d} (x,t) \  &=& q_{d}(x)  \   e^{2 a_{H} f_{d}(x)  \ t };
\label{t-GPDs-H}
\ea
\ba
{\cal{E}}_{u} (x,t) \  &=& q_{u}(x) (1-x)^{\gamma_{u}} \   e^{2 a_{E}  \  f_{u}(x) t }; \ \ \   \\ \nonumber
{\cal{E}}_{d} (x,t) \  &=& q_{d}(x)  (1-x)^{\gamma_{d}} \   e^{2 a_{E} f_{d}(x) t },
\label{t-GPDs-E}
\ea
 where
 $$ f_{u}(x) =  \frac{(1-x)^{2+\epsilon_{u}}}{(x_{0}+x)^{m}},$$
 $$f_{d}(x) = (1+\epsilon_{0}) \frac{(1-x)^{2+\epsilon_{d}}}{(x_{0}+x)^{m}}. $$
 with $a_{H}, \gamma, \epsilon_{i}, x_{0}, m$ being
   the parameters determined from the analysis
  of the  existing  experimental data of the electomagnetic form factors.

    On the basis of our GPDs with PDFs
    ABM12 \cite{ABM12},     we calculated the hadron form factors
     by  numerical integration
   and then
    by fitting these integral results by the standard dipole form with some additional parameters
\ba
  F_{1}(t)  &=& (4m_p - \mu t)/(4m_p -  t ) \\ \nonumber
     && 1/(1 + q/a_{1}+q^{2}/a_{2}^2 +  q^3/a_{3}^3)^2
 \label{Gt}
\ea
 where $m_p$ is the  proton mass and $a_i$ are free parameters.
    That is slightly  different from
  the standard dipole form of two additional terms with small sizes of the coefficients.
  The matter form factor 
\ba
 A(t)= && \int^{1}_{0} x \ dx  \\	\nonumber
 && [ q_{u}(x)e^{2 \alpha_{H} f)_{u}(x \ t  }   
  + q_{d}(x)e^{ 2 \alpha_{H} f_{d}(x)  \ t}  ]
\ea
 is fitted   by the simple dipole form
  $$  A(t)  =  \Lambda^4/(\Lambda^2 -t)^2 $$
  where $\Lambda$ is a free parameter,
  which equal $1.6$ GeV$^2$ for the proton case.
        These form factors will be used in our model of the proton-proton and proton-antiproton elastic scattering
        and  further in one of the vertices of  pion-nucleon elastic scattering.

\section{Compton and magnetic transition form factors }

        It is very important to check the obtained $t$-dependence of GPDs,
 as it  determined the $t$-dependence of the gravitomagnetic form factor of nucleons, which in turn impact on momentum transfer dependence of the differential cross sections.
     Let us calculate the moments of  GPDs with inverse power of $x$.  This  gives us the
     Compton form factors $R_{V}((t)$, $R_{T}(t)$.
   Using the obtained form factors, the reaction of the real Compton scattering
    can be calculated \cite{CompFF}.
 For  $H^{q}(x,t)$,  $E^{q}(x,t)$ with  PDFs  from the
  work \cite{Khang-16}, 
   which was chosen on the basis of the analysis carried out in \cite{GPD-PRD14}
  and with the parameters
  obtained in our fitting procedure of  describing the proton and neutron electromagnetic
   form factors   in \cite{GPD-PRD14}.
   The form factors $R_i$ are determined
      \ba
 R_{i}(t) =  \sum_{q} e^{2}_{q} \int_{0}^{1} \frac{dx}{x} {\cal{F}}j_{q}(x,\xi=0,t),
\ea
 where ${\cal{F}}j_{q}$ are equal to $H_{q}$, $E_{q}$ and $\tilde{H}_{q}$ and give the form factors $R_{V}(t)$, $R_{T}(t)$, $R_{A}(t)$,
 respectively.  

  In the present work  for $\tilde{H}^{q}(x,t)$ we take $\Delta q$ in the form \cite{Khang-16}
 for  NNLO $Q_0=2$ GeV$^2 $
\ba
 x \Delta_{q}(x,Q_{0}) = N_{q} \eta_{q} x^{a_{q} } (1-x)^{b_{q} } (1 + c_{q} x).
\ea
 Assuming $SU(3)$ flavor symmetry of $\Delta \bar{q} $,  the coefficient $N_{q}$ is determined as
\ba
 \frac{1}{N_{q}} = (1+c_{q} \frac{a_{q}}{1+a_{q}+b_{q}} ) \ B(a_{q},b_{q}+1),
\ea
 where $\ B(a_{q},b_{q}+1)$ is determined by
 \ba
 B(a,b)= \frac{\Gamma(a) \Gamma(b)}{\Gamma(a+b)} = \int_{0}^{1} t^{a-1} (1-t)^{b-1} dt.
 \ea
          The results of our calculations of the Compton form factors are shown in Fig. 1(a,b).
          The form factors $R_{V}(t)$ and $R_{T}(t)$ have a similar momentum transfer dependence but essentially differ in size.
  On the contrary, the axial form factor $R_{A}$ has an essentially different $t$ dependence.
     The calculations of $R_{i}$  on the whole, correspond to the calculations of  \cite{DK-13}.

     The differential cross section
      of the real Compton scattering
       can be written as   \cite{DK-13} 
  \ba
  \frac{d\sigma}{dt} =&&  \frac{\pi \alpha^{2}_{em}}{s^{2}} \frac{(s-u)^{2}}{-u s} \\ \nonumber
 && [R_{V}^{2}(t) \ - \ \frac{t}{4 m^{2}} R^{2}_{T}(t) 
    + \frac{t^{2}}{(s-u)^{2}} R^{2}_{A} (t)],
    \label{RCS}
\ea
  where $R_{V}((t)$, $R_{T}(t)$, $R_{A}(t)$ are the form factors given by the $1/x$
  moments of the corresponding GPDs $H^{q}(x,t)$,  $E^{q}(x,t)$, $\tilde{H}^{q}(x,t)$ .


\begin{figure}
\vspace{-1.cm}
\includegraphics[width=.45\textwidth]{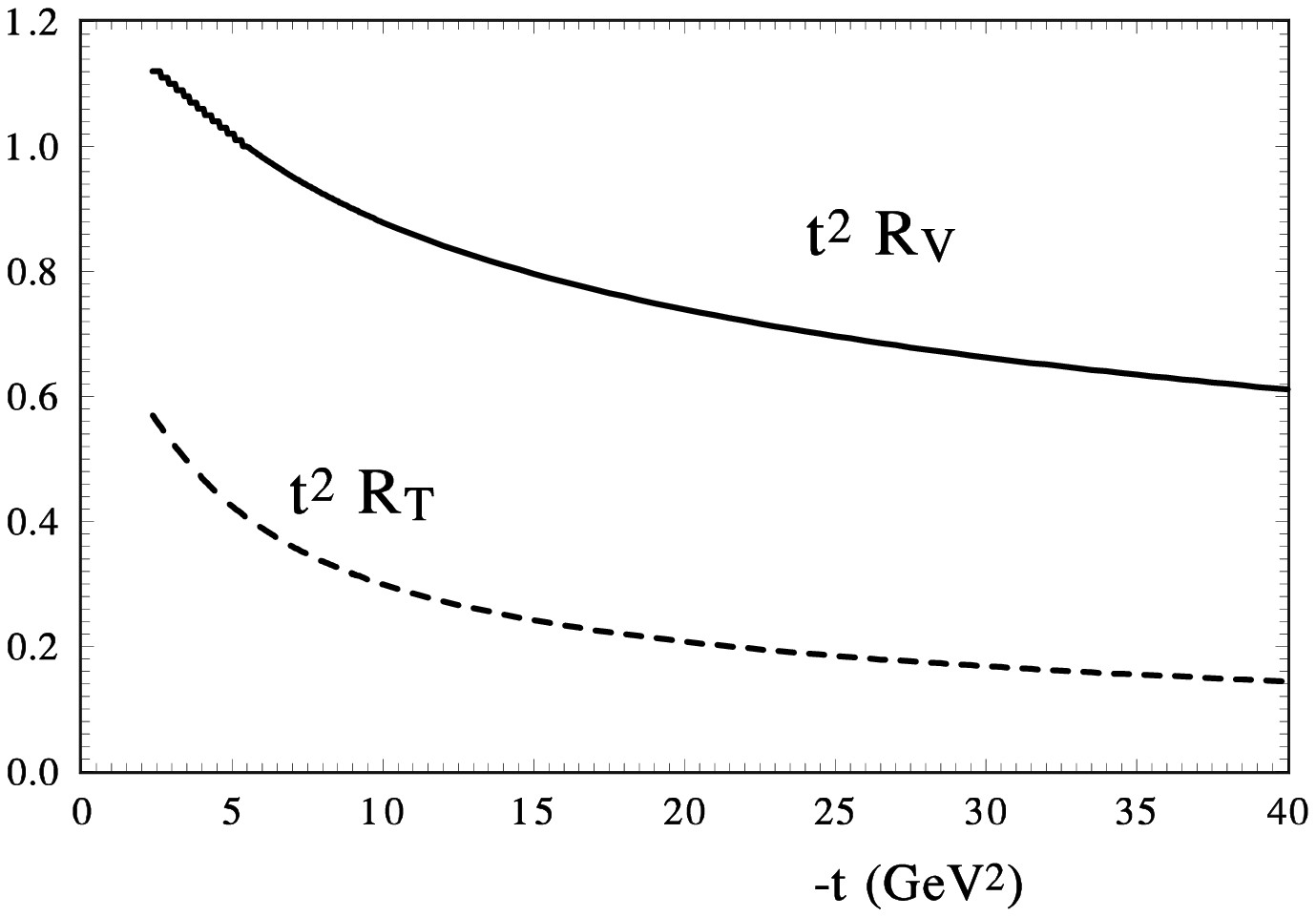} 
\includegraphics[width=.45\textwidth]{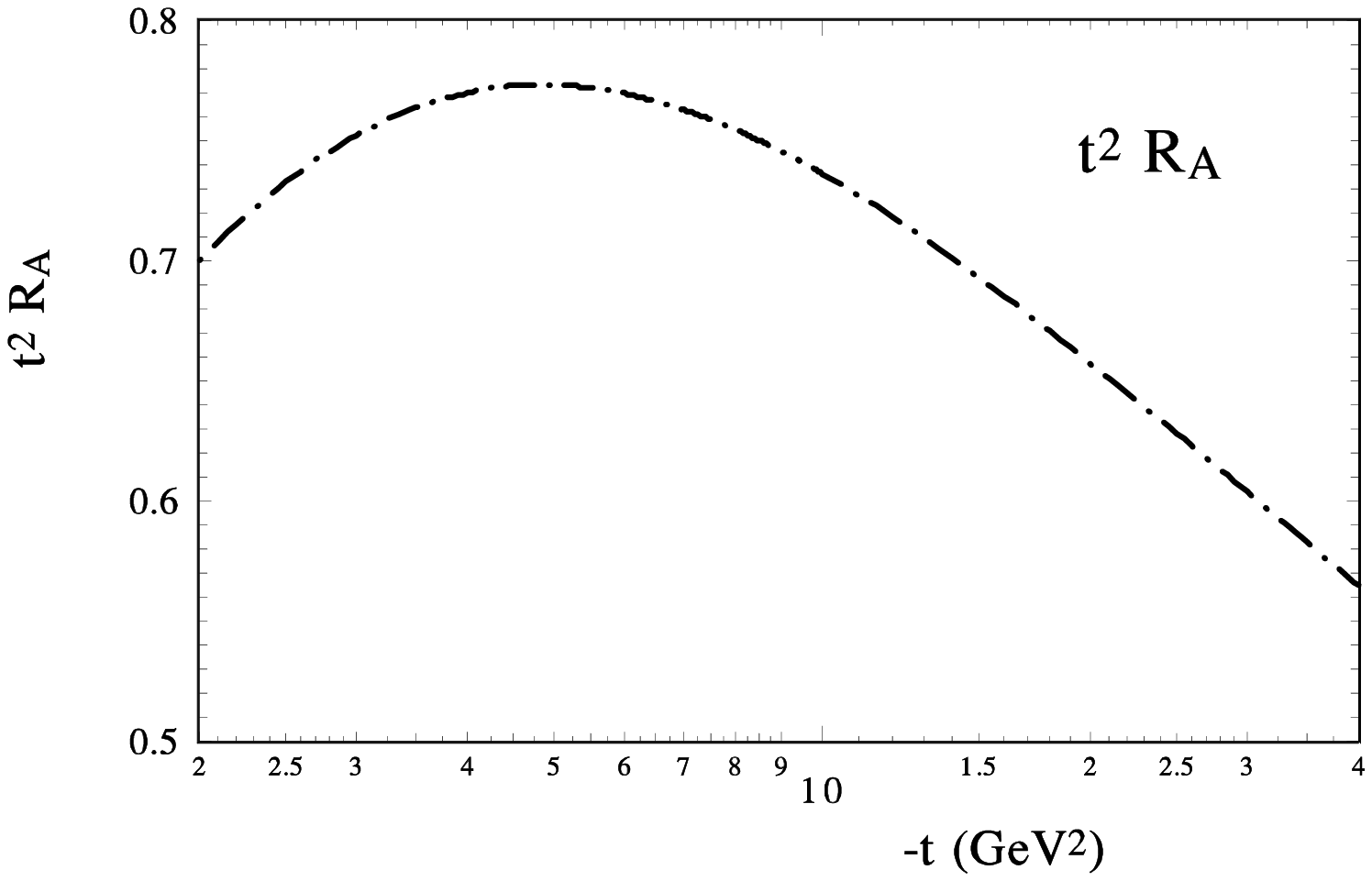} 
\vspace{0.5cm}
\vspace{1.cm}
\caption{  
 The Compton form factors 
  a) [top] $t^2 R_{V}(t)$   and $t^2 R_{T}(t)$;
   b)[bottom]   , 
 $t^2 R_{A}(t)$.
  }
\label{Fig_1b}
\end{figure}

        The results for the cross sections are presented in Fig.2. Except for  very   large angles at low energies
        the coincidence with  experimental data is sufficiently good.

\begin{figure}
\vspace{-1.cm}
\begin{center}
\includegraphics[width=.45\textwidth]{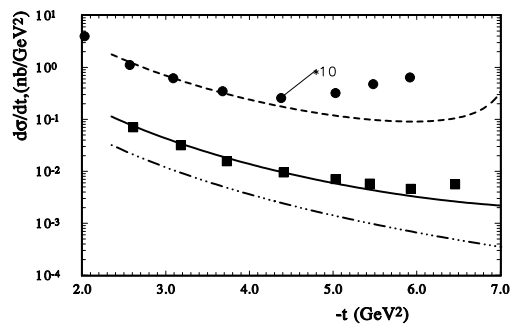} 
\vspace{0.5cm}
\caption{
The differential Compton cross sections 
 (the lines are our calculations at
$s=8.9$ GeV$^2$,
  $s=10.92$ GeV$^2$  
 and $s=20$ GeV$^2$,
   the data points are for $s=8.9$ GeV$^2$ (circles)\cite{dt-Cmpt};
  $s=10.92$ GeV$^2$ (squares)  \cite{dt-Cmpt}. 
  }
 \end{center}
\label{Fig_3}
\end{figure}

  To check the obtained
   momentum dependence of the spin-dependent part of GPDs $ E_{u,d}(x,\xi=0,t) $,
   we can calculate the magnetic transition
  form factor which is determined by the difference of $ E_{u}(x,\xi=0,t) $ and $ E_{d}(x,\xi=0,t) $.
 For the magnetic $N \rightarrow \Delta $ transition form factor $G^{*}_{M}(t)$,  in the large $N_{c}$  limit,
the relevant $GPD_{N\Delta}$ can be expressed in terms of the isovector GPD
    yielding the sum rule \cite{Guidal}
\ba
  G^{*}_{M} (t) =  &&  \frac{G^{*}_{M} (t=0)}{k_{v} }     \nonumber \\
  && \int_{-1}^{1} dx ( E_{u}(x,\xi,t) - E_{d}(x,\xi,t) )
\ea
  where $k_{v}=k_{p} - k_{n} =3.70 $.


\begin{figure}
\vspace{-1.cm}
\includegraphics[width=.45\textwidth]{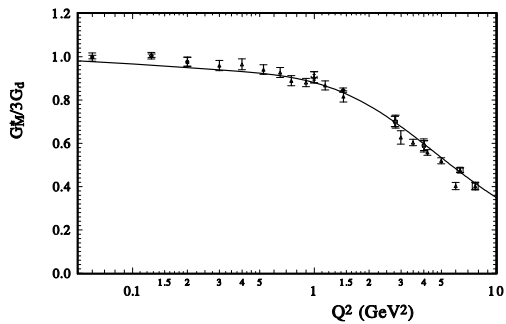} 
\vspace{0.5cm}
\caption{
The transition magnetic  form factors  $G^{*}_{M}(Q^2)/(3G_{d})$
(line- our calculations, points are the experimental data \cite{ff-Gtr}).
  }
\label{Fig_1}
\end{figure}

  The results of our calculations, based on eqs. (2) and (3),  are presented in Fig.3.
 The  experimental data  exist up to
   $-t =8 $ GeV$^2$ and our results  show a sufficiently good coincidence with experimental data.
   It is confirmed that the form of the momentum transfer dependence of  $E(x,\xi,t)$ determined in our model  is correct.

\section{Hadron form factors and elastic nucleon-nucleon scattering}

\begin{figure}
\begin{center}
\includegraphics[width=.49\textwidth]{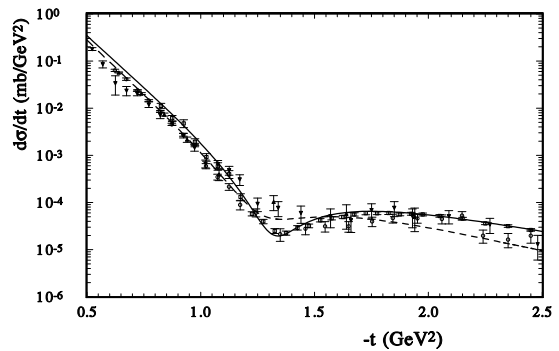} 
\includegraphics[width=.49\textwidth]{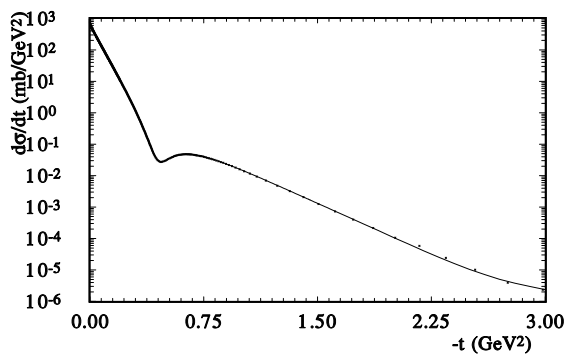} 
\vspace{1.cm}
\caption{ a)
[top] The HEGS model calculations of the differential cross sections of elastic scattering of proton-proton (hard line)
  and proton-antiproton (dashed line) at $\sqrt{s} = 52.2 $ GeV,
  ( circles, quires, triangles up and triangles down -
  \cite{Rub19am,Adam19,Schiz19,Aker19,Cul19,Rub19,Bri19}) \\
  \hspace{4.cm} b) [bottom] $pp$  elastic cross sections
  at $\sqrt{s} = 13 $ TeV (line - the HEGS model calculations, points - the data \cite{T66,T67}).
  }
 \end{center}
\label{Fig_4}
\end{figure}

   In the framework of the high energy generalized structure   (HEGS)
        model  of  elastic nucleon-nucleon scattering
             both hadron  electromagnetic and gravitomagnetic form factors were used.
        This allows us to build a model with a minimum number of fitting
        parameters \cite{HEGS0,HEGS1,NP-hP}.

   The Born term of the elastic hadron amplitude can now be written as
  \begin{eqnarray}
F_{h}^{Born}(s,t)&&=h_1  F_{1}^{2}(t) F_{a}(s,t)  (1+r_1/\hat{s}^{0.5})
    \\ \nonumber
     + && h_{2} \  A^{2}(t) \ F_{b}(s,t) \     \\
      \pm  &&  h_{odd} \  A^{2}(t)F_{b}(s,t)\ (1+r_2/\hat{s}^{0.5}),  \nonumber
    \label{FB}
\end{eqnarray}
  where $F_{1}(t)$ is the electromagnetic proton form factor, which represents charge distribution in the proton,  and $A(t)$ is the gravitation form factor which represents the matter distribution in the proton;
 hence,  both (electromagnetic and gravitomagnetic) form factors are used.
 The parameters are determined in \cite{HEGS1}
  where $F_{a}(s,t)$ and $F_{b}(s,t)$  have the standard Regge form: 
  \begin{eqnarray}
 F_{a}(s,t) \ = \hat{s}^{\epsilon} \ e^{B(\hat{s}) \ t}; \  
 F_{b}(s,t) \ = \hat{s}^{\epsilon} \ e^{B(\hat{s})/4 \ t},
\end{eqnarray}
 where $   \hat{s}=s \ e^{-i \pi/2}/s_{0}$ ;  $s_{0}=4 m_{p}^{2} \ {\rm (GeV^2)}$, and
  $h_{odd} = i h_{3} t/(1-r_{0}^{2} t) $.
  The intercept $1+\epsilon =1.11$ was chosen from the data of
    different reactions and was fixed by the same size for all terms of 
  the scattering amplitude.
 The slope of the scattering amplitude has the standard logarithmic dependence on the energy
 $   B(s) = \alpha^{\prime} \ ln(\hat{s}) $
  with $\alpha^{\prime}=0.24$ GeV$^{-2}$  and with some small additional term \cite{HEGS1},
    which reflects the small non-linear behavior of  $\alpha^{\prime}$ 
     \cite{Sel-Df16}.
The final elastic  hadron scattering amplitude is obtained after unitarization of the  Born term
 by the standard eikonal representation.
  The model is very simple from the viewpoint of the number of fitting parameters and functions.
  There are no any artificial functions or any cuts which bound the separate
  parts of the amplitude by some region of momentum transfer.

        In the framework of the model, the description of
         experimental data was obtained simultaneously
        at the large momentum transfer and in the Coulomb-hadron region, using the CNI phase \cite{selmp1,Selphase},
        in the energy range from $\sqrt{s}=9 $ GeV
        up to LHC energies.
In the basic form of the HEGS model $3416$ experimental points were included in our analysis
 in the energy region   $9.8$ GeV $\leq \sqrt{s} \leq 8. $ TeV
 and in the region of momentum transfer $0.000375 \leq |t| \leq 15 $ GeV$^2$.
 The experimental data of  proton-proton and proton-antiproton elastic scattering are included
 in 92 separate sets of 32 experiments, 
 including recent data of the TOTEM Collaboration
 at $\sqrt{s}=8$ TeV.  
  The whole Coulomb-hadron interference region,
  where the experimental errors are remarkably small,
    was included in our examination of  experimental data.
 Our model of the GPDs leads to a good description of the proton and neutron  electromagnetic form factors and their elastic scattering simultaneously.
   It allows one to find some new features in the differential cross section of $pp$-scattering in the
   unique experimental data  of the TOTEM collaboration at $ \sqrt{s}=13 $ TeV (small oscillations
        \cite{Sel-PL19}           
   and anomalous behavior at small momentum transfer \cite{anom13-20} ).
   The inclusion of the spin-flip parts of the scattering amplitude allows one to describe the low energy  experimental polarization data of the $pp$ elastic scattering  \cite{Symmetry},
   which are shown in the corresponding figures in \cite{Symmetry}.


         Figure 4(a) represents the description at  $\sqrt{s}=52.8 $ GeV and Figure 4(b)
  shows the model calculations for $\sqrt{s}=13 $ TeV, which coincide well with
  the recent experimental data of the TOTEM Collaboration \cite{T66,T67}.

\section{GPDs of pion }

\begin{figure}
\begin{center}
\includegraphics[width=.49\textwidth]{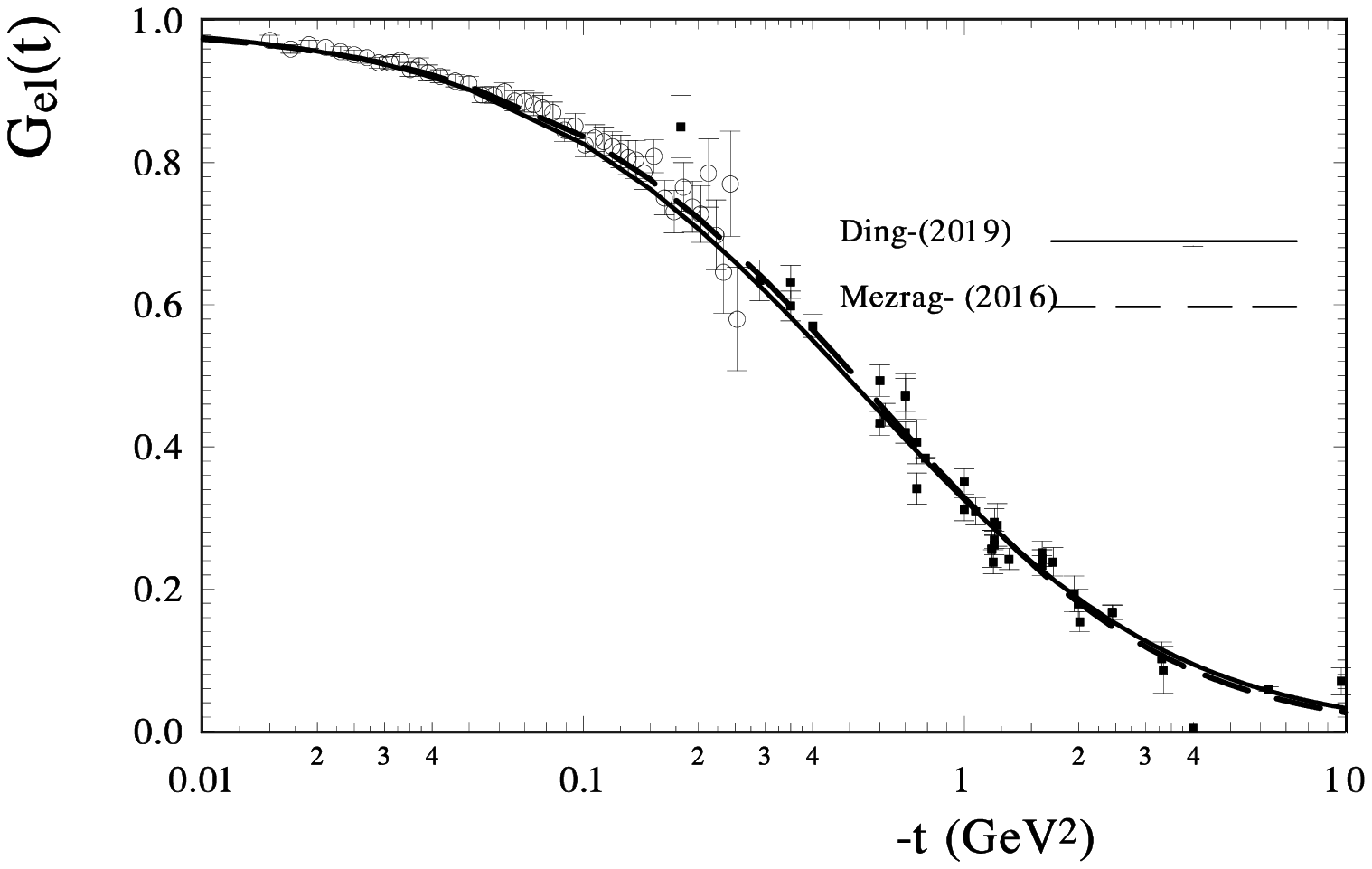} 
\includegraphics[width=.49\textwidth]{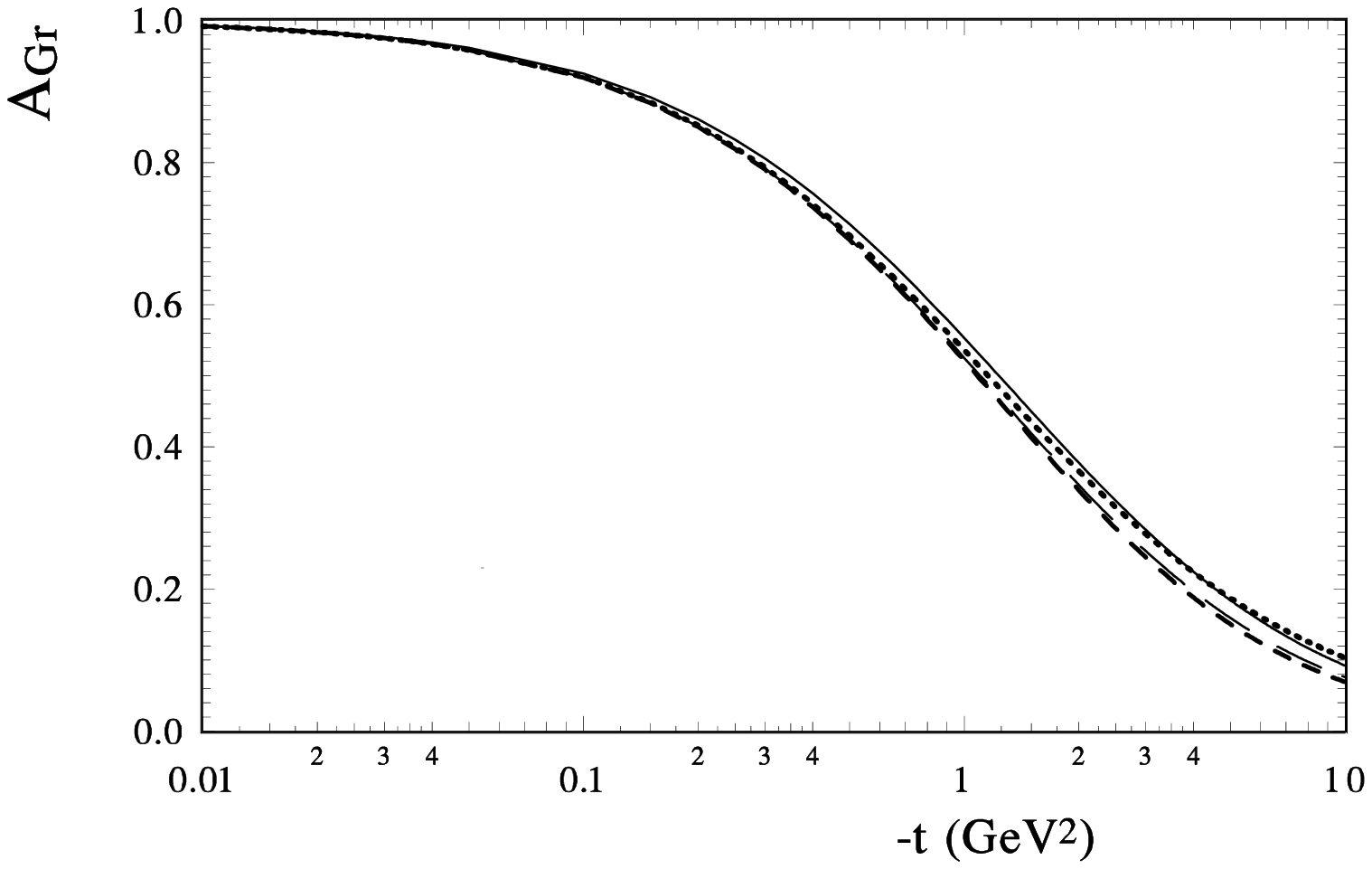} 
\vspace{1.cm}
\caption{a) [top] The electromagnetic form factor of the $\pi$-meson
(hard and dashed curves - our calculations with  PDF \cite{Mez16} and  and \cite{RayT}, respectively;
 the circles and squares - the experimental data \cite{ff1,ff2,ff3,ff4,ff5,ff6}) \\
 b) [bottom] the gravitomagnetic form factor of the pion
 with the normalization $A_{Gr}(t=0)=1$
 (the hard and dashed curves -
 our calculation with the PDF \cite{Mez16} and   \cite{RayT}, respectively);
 long-dashed and tiny-dashed curves - the fits of our integral calculations by a simple monopole form.
}
\end{center}
\label{Fig_5}
\end{figure}

      The pion structure in some sense is simpler than the nucleon structure.
 In the nucleon there are 3 constituent quarks that can create  different configurations, for example,
  such as "Mercedes star" or a linear structure
  with a quark at one end and a di-quark at the other.
  These configurations can lead to   different results for hadron interactions,
  for example, the Odderon-hadron coupling.
  For a meson we have only two quark states
$$ |\pi> =|q\bar{q}> + |q\bar{q} \ q\bar{q}> + |q\bar{q} \ g> .....   .$$
  It is needed to note that the standard definition of the pion form factor
  through the matrix elements of the electromagnetic vector current
 \ba
 V_{\mu}(x) =  
  e_{u} \bar{u}(x)\gamma_{\mu}u(x) -   
  e_{d}  \bar{d}(x)\gamma_{\mu}d(x),
\ea
  gives
\ba
 \langle   \pi^{+}(\vec{p^{'}}) |V_{\mu}(0)| \pi^{+}(\vec{p})\rangle
  = (p_{\mu}^{'} + p_{\mu}) F_{\pi}(Q^{2}),
\ea
  with $Q^{2}=-q^2$ and $F_{\pi}(Q^{2})$ being the space-like form factor of the pion  \cite{ETM}.
  It is related with the separate quark contributions
   \ba
   F_{\pi}(t) = e^{u} \ F_{\pi}^{u} - e^{d} \ F_{\pi}^{d}.
\ea
  For the definition of the electromagnetic form factor of the pion there are
  many different approximations beginning with the standard monopole form
  \ba
  F_{\pi}=\Lambda^2/(\Lambda^2 -t),
  \ea
  (with $\Lambda$ as a free parameter determined from experimental data),
   including the Regge exponential form
   $$  F_{\pi}=y^{-\alpha_{\pi}(t)}   \frac{e^{t-m_{\pi}^{2}}}{\Lambda^2} $$
  and  monopole form with polynomial form of $t$ dependence \cite{Mez-14}
$$  F_{\pi} (t=M^{2}z)=\frac{1}{1+0.44 z +0.06 z^{2}+0.00033 z^{3}} . $$
and  in complicated form of $t$ dependence \cite{Meln03}
    $$
    F_{\pi}(Q^2) = \frac{1}{1+Q^2/m_{\rho}^{2}}(\frac{1+ c_{1} Z +  c_{2} Z^{2}}{1+ c_{1} Z + c_{2} Z^{2} + c_{3} Z^3}),
    $$
    where $Z=Log(1+Q^2/\Lambda^2)$ and $\Lambda$ is the QCD scale parameter.
    Such a form is similar
to that proposed in \cite{Watanabe}  within a dispersion relation analysis; however, the presented form uses two additional parameters and takes a rather large value of $ \Lambda = 1$ GeV.

  For the pion Generalized parton distribution we have the standard definition through the  matrix element, for example \cite{Mez-14}
\ba
&& H^{q}_{\pi}(x,t,\xi)  =\frac{1}{2} \int \frac{dz^{-}}{2\pi} e^{i x P^{+} z^{-} }   \\
&& \langle \pi,P+\frac{\Delta}{2}|\bar{q}(-\frac{z}{2}) \gamma^{+}[-\frac{z}{2};\frac{z}{2}] q(\frac{z}{2})|\pi,P-\frac{\Delta}{2}\rangle_{z^{+}=0,x_{\perp}=0},     \nonumber
\ea
 with the skewness $\xi=-\Delta^{+}/(2P^{+})$ and the momentum transfer $t=-\Delta^{2}$. 
  Taking into account the charge conjugation  corresponding  to separate
   quarks of GPDs, we obtain
\ba
 H^{u}_{\pi^{+}}(x,t,\xi=0) = -  H^{d}_{\pi^{+}}(-x,t,\xi=0)
\ea
 and  for the charged pions
 \ba
 H^{u}_{\pi^{\pm}}(x,t) = H^{d}_{\pi^{\pm}}(x,t).
 \ea
For the full form of pion GPDs we take the same ansatz as we used for the nucleon case.
 We have focused on the zero-skewness limit, where  GPDs
have a probability-density interpretation in the longitudinal Bjorken x and
the transverse impact-parameter distributions.
The pion form factors will be obtained  by integration  over $x$ in the whole range $0 - 1$.
 Hence, the obtained  form  factors will be dependent  on the  
   forms of PDF  at the ends of the integration region.
  Some  PDFs  have the polynomial form of $x$ with
     different power.  Some others have the exponential dependence of $x$.
  As a result, the behavior of  PDFs, when $x \rightarrow 0$ or $x \rightarrow 1$,  can  impact  the
    form of the calculated form factors.

\begin{figure}
\begin{center}
\includegraphics[width=.49\textwidth]{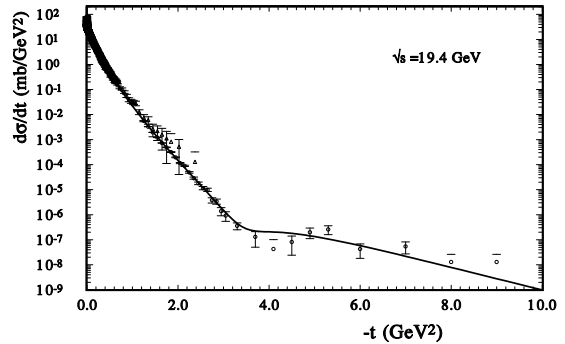} 
\includegraphics[width=.49\textwidth]{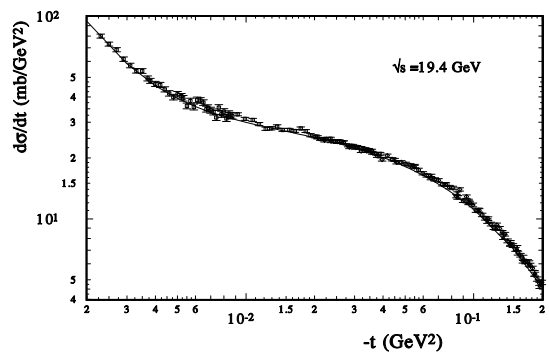} 
\vspace{1.cm}
\caption{ 
 The $d\sigma/dt$ of the $\pi^{-} p$ elastic scattering
  at $\sqrt{s}=19.4$ GeV ([top] full region of examined $t$
  (the corresponding part of the total $\chi^2$ given  these data is
  $\chi^2/n = 442/294 = 1.5$)
   and [bottom]
  small region of $t$
  (the corresponding part of the total $\chi^2$ given these data is
  $\chi^2/n = 166/132 = 1.26$).
  [On these figures and others the comparison
    of the experimental data with theoretical calculations is shown with additional normalization
    coefficient equal to unity and with only statistical experimental errors]
 the circles and squares - the experimental data \cite{ff1,ff2,ff3,ff4,ff5}).
}
\end{center}
\label{Fig_6}
\end{figure}

\begin{figure}
\begin{center}
\includegraphics[width=.49\textwidth]{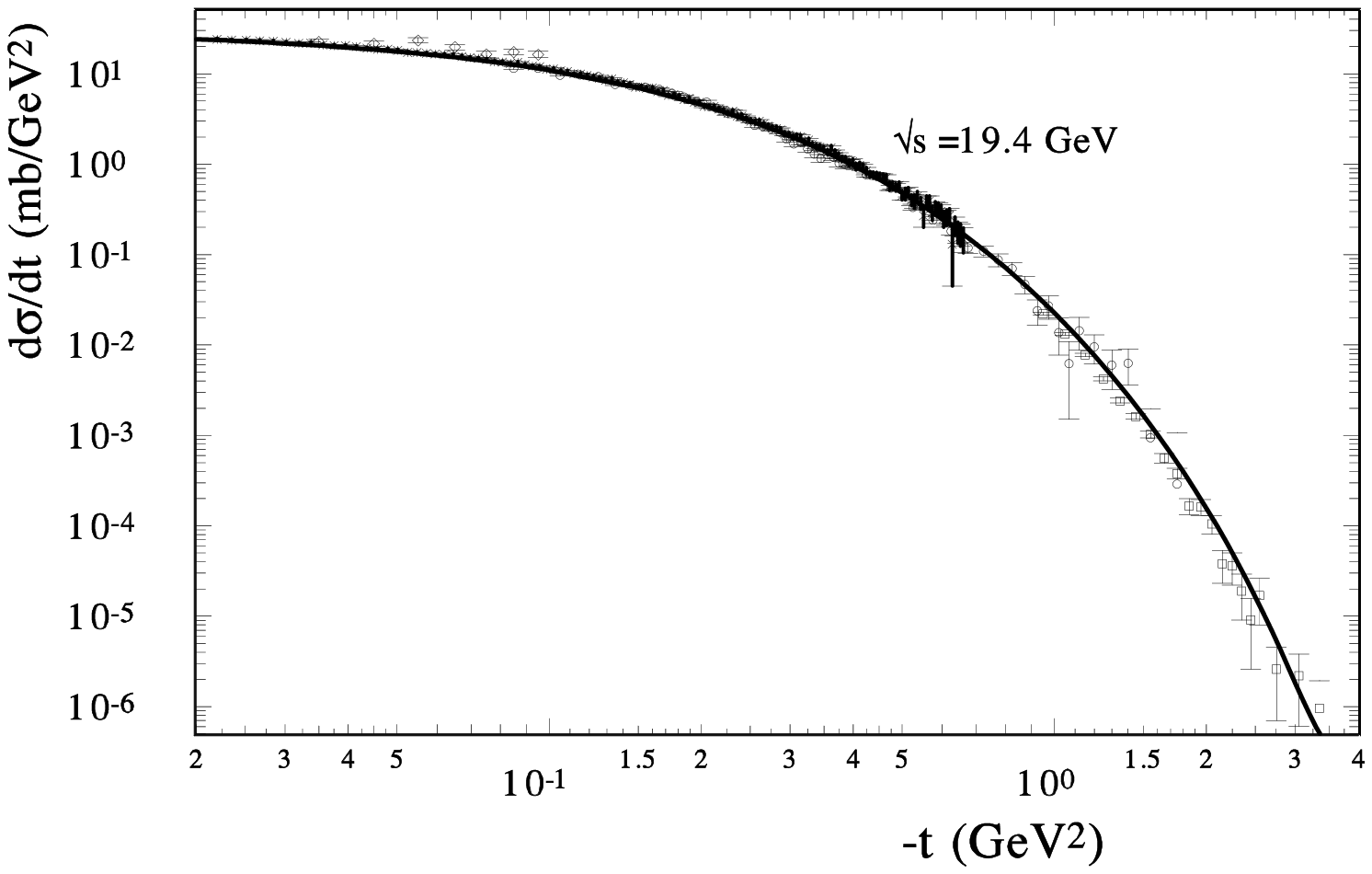} 
\includegraphics[width=.49\textwidth]{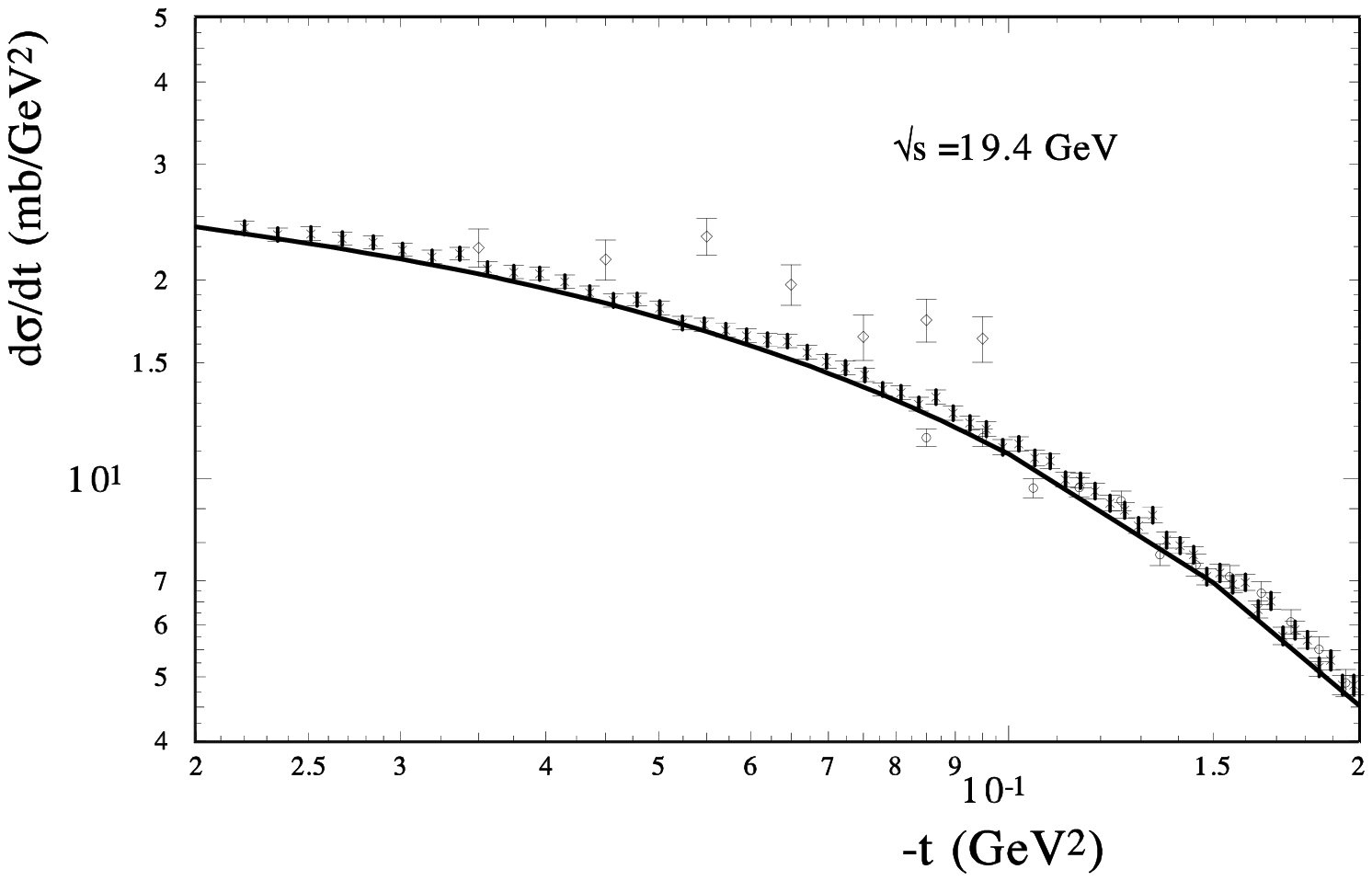} 
\vspace{1.cm}
\caption{a) The $d\sigma/dt$ of the $\pi^{+} p$ elastic scattering
  at $\sqrt{s}=19.4$ GeV ([top] full region of examined $t$
  (the corresponding part of the total $\chi^2$ given these data is
  $\chi^2/n = 198/212 = 0.93$)
  and [bottom]
  small region of $t$
  (the corresponding part of the total $\chi^2$ given these data is
  $\chi^2/n = 79.2/79 = 1.$).
 the circles and squares - the experimental data \cite{ff1,ff2,ff3,ff4,ff5}).
}
\end{center}
\label{Fig_7}
\end{figure}

  Various Collaborations have determined the  PDF sets  from  inelastic processes only in  some region of $x$, which are further
 approximated to $x=0$ and $x=1$.
  Also, there is a serious problem in determining  the main
  ingredient of GPDs of a pion - the basic form of  parton distribution functions.
  The predictions based on the perturbative QCD and
the calculations using  different approaches support the pdf in the form $(1 - x)^2$
as $x \rightarrow 1$ (see for example \cite{Hecht} and complicated analysis carried out in \cite{Roberts}
   However, the constituent quark model and calculation in the framework of
   the Nambu-Jona-Lasino model lead to  linear behavior $(1-x)_{x \rightarrow 1}$.
   Several next-to-leading order (NLO) analyses of the Drell-Yan data show
   that the valence distribution turned out to be rather hard at high momentum fraction x ,
typically showing  only a linear or slightly faster falloff.
  Correspondingly, there are many different forms of the PDF of a pion.
    For example,  
      \cite{Wat-16,Wat-18}
 \ba
  \nu^{\pi}_{bare}(x, Q^{2}_{0}) = A_{0} x^{\alpha} \ (1 \ - x) ^{\beta} \nonumber
 \ea
 with $\alpha=1.8; \ \beta=1.8$ ; \\
  or (M. Aicher et al. (2010)) \cite{Aich10}
  \ba
   \nu^{\pi}_{bare}(x, Q^{2}_{0}) = N_{\nu} x^{\alpha} \ (1 \ - x) ^{\beta} (1+ \gamma x^{2}) \nonumber
  \ea
 with $\alpha=1.06; \ \beta=1.75; \ \gamma= 1.4$. We examine many of them   \cite{Mez16,Han18,Dan19,Bour-20}
     and keep two PDFs leading to approximately  the same results
  and giving the good description of the existence experimental data of pion form factor:
   one is (L. Chang \cite{Mez16} )
  \ba
  \nu^{\pi}(x)  \ = N \ 3.47 \ x^{0.021}   \ (1 - x)^{2.33} ;
  \ea
and    R. Sufian    \cite{Sufian-20}
     \ba
  \nu^{\pi}(x) = N \frac{ x^{-\alpha} \ (1 \ - x) ^{\beta} (1+\gamma x)}{(B(\alpha+1,\beta+1)- \gamma B(\alpha+2),\beta+1)}
  \ea
  where $B(\alpha,\beta)$ is the incomplete Gamma function.
  There are two variants: with $\gamma=0$ and with $\gamma=4.28$.

 In first variant   $ \alpha = -0.17$,  $ \beta = 1.24$  and
 in the second variant     $ \alpha = -0.22$,  $\beta = 2.12$.
  In that work it was noted that both variants give practically the same result.

  In our fitting procedure with variation of the slope parameters of the GPDs
  both variants give close values for the  constants of the electromagnetic and gravitomagnetic form factors.
  In the first case $\Lambda^{2}_{em} = 0.49 \pm 0.04$ and  $\Lambda^{2}_{gr} = 1.12 \pm 0.15 $,
  and in the second case  $\Lambda^{2}_{em} =  0.47 \pm 0.08$ and  $\Lambda^{2}_{gr} = 1.07 \pm 0.09 $,

 On the basis of our GPDs with
    PDFs,    we have calculated the pion form factors
     by  numerical integration
   and then
    by fitting these integral results by the standard monopole form,
    which gives the power like scaling \cite{Brodsky},
      and obtained $\Lambda^{2}_{\pi} =0.5$.
  In Fig.5a, the comparison of our calculation with the existing experimental data of the
  pion form factor is presented. It is seen that the difference between the calculations
  of  our two chosen  PDFs is small, 
 both variants give the $\chi^2$  values
that are the same within the estimated uncertainty.

   The matter form factor $A_{Gr}^{\pi}(t)$ is calculated as the second Mellin moment
\ba
 A_{Gr}^{\pi}(t)=  \int^{1}_{0} x \ dx \
 q_{\pi}(x)e^{2 \alpha_{\pi} f(x) / t  } 
\ea
 and is fitted   by the simple dipole form  $  A(t)  =  \Lambda^4/(\Lambda^2 -t)^2 $.
        These form factors will be used in our model of the $\pi^{+}p$ and $\pi^{-}p$  elastic scattering.
 In Fig.5b, our calculations of the second momentum of GPDs of a pion
  are shown. Again, we see that the impact of different PDFs is tangible only
  at large momentum transfer.

 \section{Hadron form factors and elastic pion-nucleon scattering }

\begin{figure}
\begin{center}
\includegraphics[width=.49\textwidth]{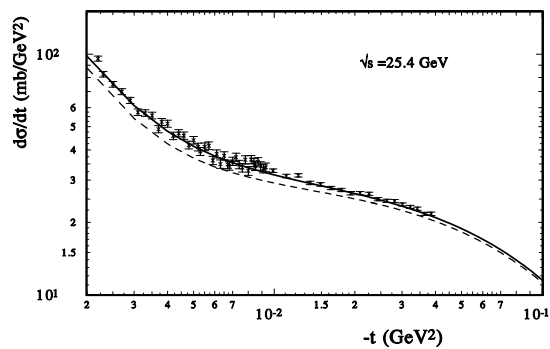} 
\vspace{0.5cm}
\vspace{1.cm}
\caption{  The differential cross sections of the  
 $\pi^{-}p$  elastic cross sections
  at $\sqrt{s} = 25.4 $ GeV (the dashed line is  the model calculations
  for the $\pi^{+}p$  elastic cross sections at this energy, points -the data \cite{25p4})
  (the corresponding part of the total $\chi^2$ given these data is
  $\chi^2/n = 69/57 = 1.2$).
  }
 \end{center}
\label{Fig_8}
\end{figure}

  Let us determine the Born terms of the elastic pion-nucleon scattering amplitude
  in the same form as we determined the elastic nucleon-nucleon scattering
   amplitudes. Using  both the (electromagnetic and gravitomagnetic) form factors
   of a pion and a nucleon, we obtain 
  \begin{eqnarray}
 F_{mh}^{Born}(s,t)=&& h_1 \ F_{1}(t)  \ F_{\pi}(t) \ F_{a}(s,t)
  \\ \nonumber
   &&  +  h_{2} \  A_{N}(t) \ A^{\pi}_{Gr}(t) \  F_{b}(s,t) \     \\  \nonumber
   && \pm R_{c}(s,t),
    \label{FB}
\end{eqnarray}
 where $F_{\pi}(t)$ is the electromagnetic pion form factor, which represents the charge distribution in the pion  and $A^{\pi}_{Gr}(t)$ is the gravitation form factor which represents the matter distribution in the pion, and
   $F_{a}(s,t)$ and $F_{b}(s,t)$  have the standard Regge form: 
   \begin{eqnarray}
 F_{a}(s,t)  = \hat{s}^{\epsilon} (1+ (1- k_{2}/(k_{1}\sqrt{\hat{s}}) ) k_{1}/\sqrt{\hat{s}}   ) e^{B(\hat{s}) \ t}; \\
  F_{b}(s,t)  = \hat{s}^{\epsilon} (1+(1+k_{2}/\sqrt{\hat{s}})/\sqrt{\hat{s}})
  e^{B(\hat{s})/4 \ t},
\end{eqnarray}
 with $   \hat{s}=s \ e^{-i \pi/2}/s_{0};  \ \ s_{0}= 1 \ {\rm GeV^2}$, 
  and
  at $t=0$ the intercept $1+\epsilon =1.11$ was chosen the same as for  nucleon-nucleon elastic
  scattering. Hence, at the asymptotic energy we have the universality
  of the energy behavior of the elastic hadron scattering amplitudes.

\begin{figure}
\begin{center}
\includegraphics[width=.45\textwidth]{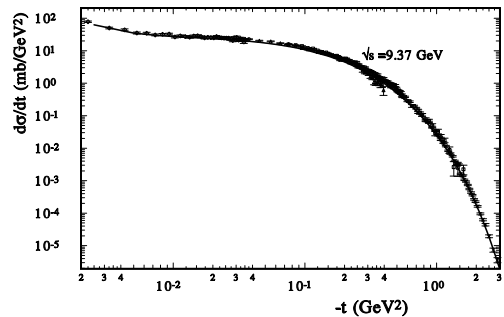} 
\includegraphics[width=.45\textwidth]{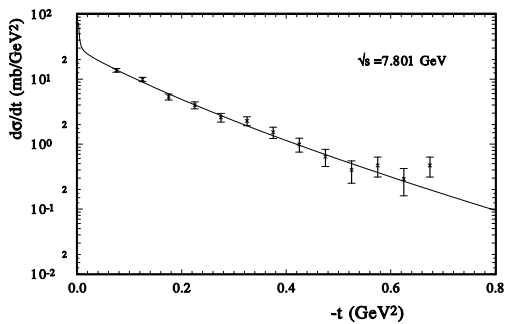} 
\vspace{0.5cm}
\vspace{1.cm}
\caption{
The differential cross sections of the elastic scattering of
 the $\pi^{-}p$  elastic cross sections
  at $\sqrt{s} = 9.73 $ GeV (the line is the model calculations, the squares and circles
  are the experimental data  \cite{derev74,apokin75,ayres76,asad84} )
   (the corresponding part of the total $\chi^2$ given these data is
  $\chi^2/n = 213/191 = 1.1$)
   and b)[bottom] the $\pi^{+}p$  elastic cross sections
  at $\sqrt{s} = 7.807 $ GeV with experimental data \cite{azinenko80}
  (the corresponding part of the total $\chi^2$ given these data is
  $\chi^2/n = 8/13 = 0.6  $).
   }
 \end{center}
\label{Fig_9}
\end{figure}

 The slope of the scattering amplitude has the standard logarithmic dependence on the energy
 $   B(s) = \alpha^{\prime} \ ln(\hat{s}) $
with $\alpha^{\prime}=0.24$ GeV$^{-2}$ (the same value as for nucleon-nucleon elastic scattering).
  Examining the pion-nucleon elastic scattering at low energies, we take into
  account the contributions of the non-leading  cross-odd Reggions using the  form factors of
  the pion and nucleon:
  \ba
 R_{c}^{Born}(s,t)=  h_{c} G_{\pi}(t) G_{N}(t) \frac{i(\pi/2\pm1)}{\sqrt{\hat{s}}} e^{b_{R} t Ln(\hat{s})}. 
    \label{FB}
\ea
with  the standard Reggion slope $b_{R}=0.9$ GeV$^{-2}$.

\begin{figure}
\begin{center}
\includegraphics[width=.45\textwidth]{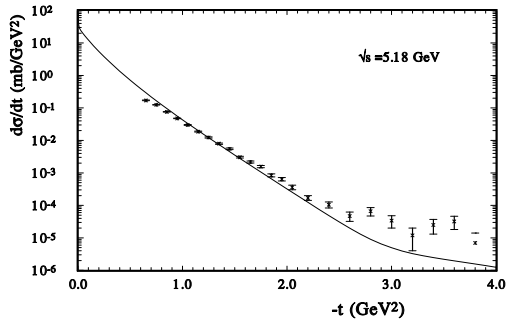} 
\vspace{0.5cm}
\vspace{1.cm}
\caption{
The differential cross sections of the elastic scattering of the
 $\pi^{+}p$  elastic cross sections
  at $\sqrt{s} = 5.18 $ GeV (the line is the model predictions, the points
  are the experimental data \cite{rub73} ). }
 \end{center}
\label{Fig_10}
\end{figure}

 As a result, only $5$ constants of interaction are included in the fitting procedure.
 The energy dependence,  the momentum transfer  dependence and the real part of the scattering amplitude
  are determined by
 the complex $\hat{s}$ and intercept. Their values do not
  change  in the fitting procedure.
The final elastic  hadron scattering amplitude is obtained after unitarization of the  Born term.
    So, at first, we have to calculate the eikonal phase
  \begin{eqnarray}
 \chi(s,b) \   =  -\frac{1}{2 \pi}
   \ \int \ d^2 q \ e^{i \vec{b} \cdot \vec{q} } \  F^{\rm Born}_{h}(s,q^2)
\end{eqnarray}
  and then obtain the final hadron scattering amplitude.
    \begin{eqnarray}
 F_{h}(s,t) = i s
    \ \int \ b \ J_{0}(b q)  \ \Gamma(s,b)   \ d b\, ; \\
   {\rm with }  \ \ \ \   \Gamma(s,b)  = 1- \exp[ \chi(s,b)] .
 \label{overlap}
\end{eqnarray}

\begin{figure}
\begin{center}
\includegraphics[width=.49\textwidth]{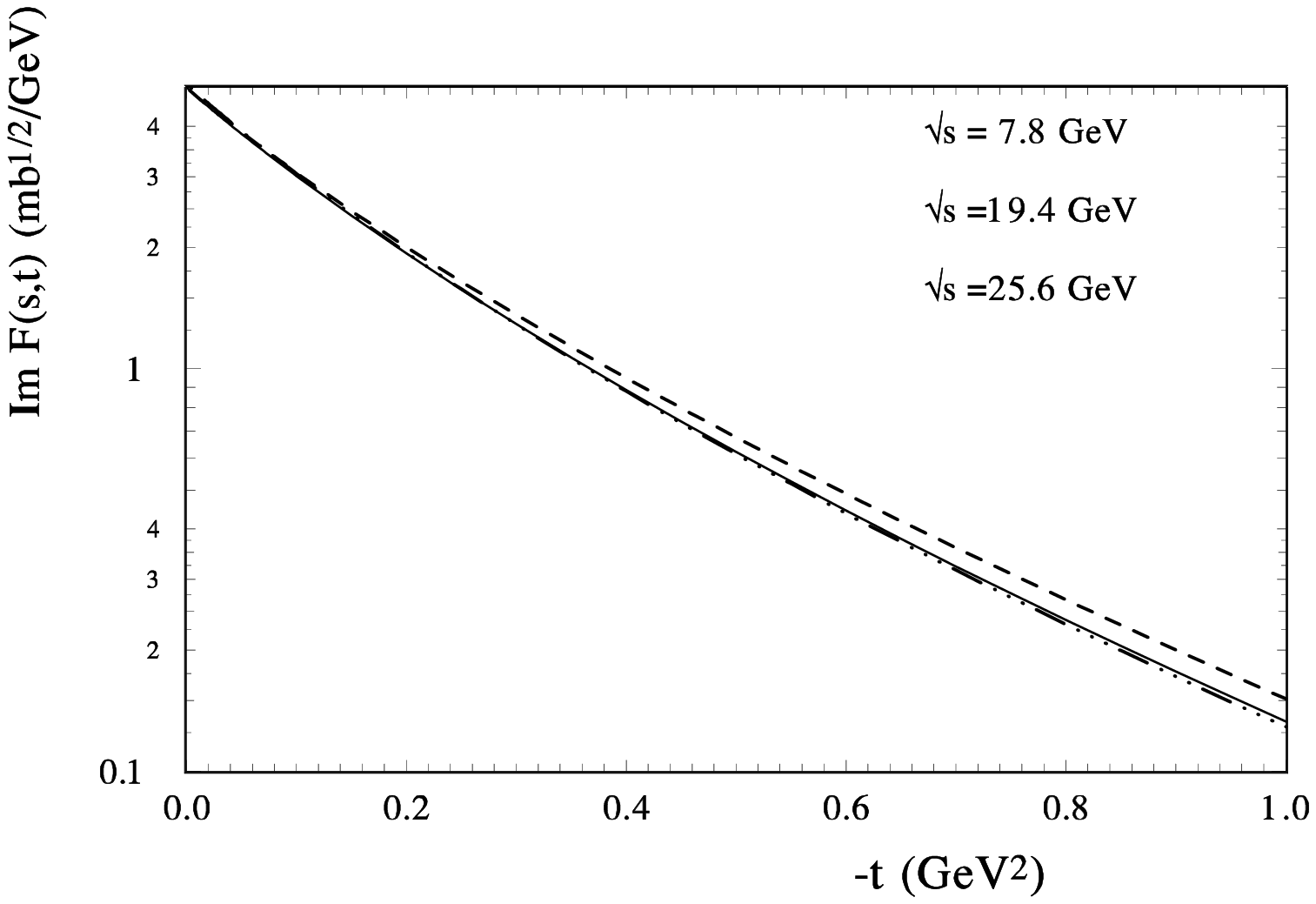} 
\includegraphics[width=.49\textwidth]{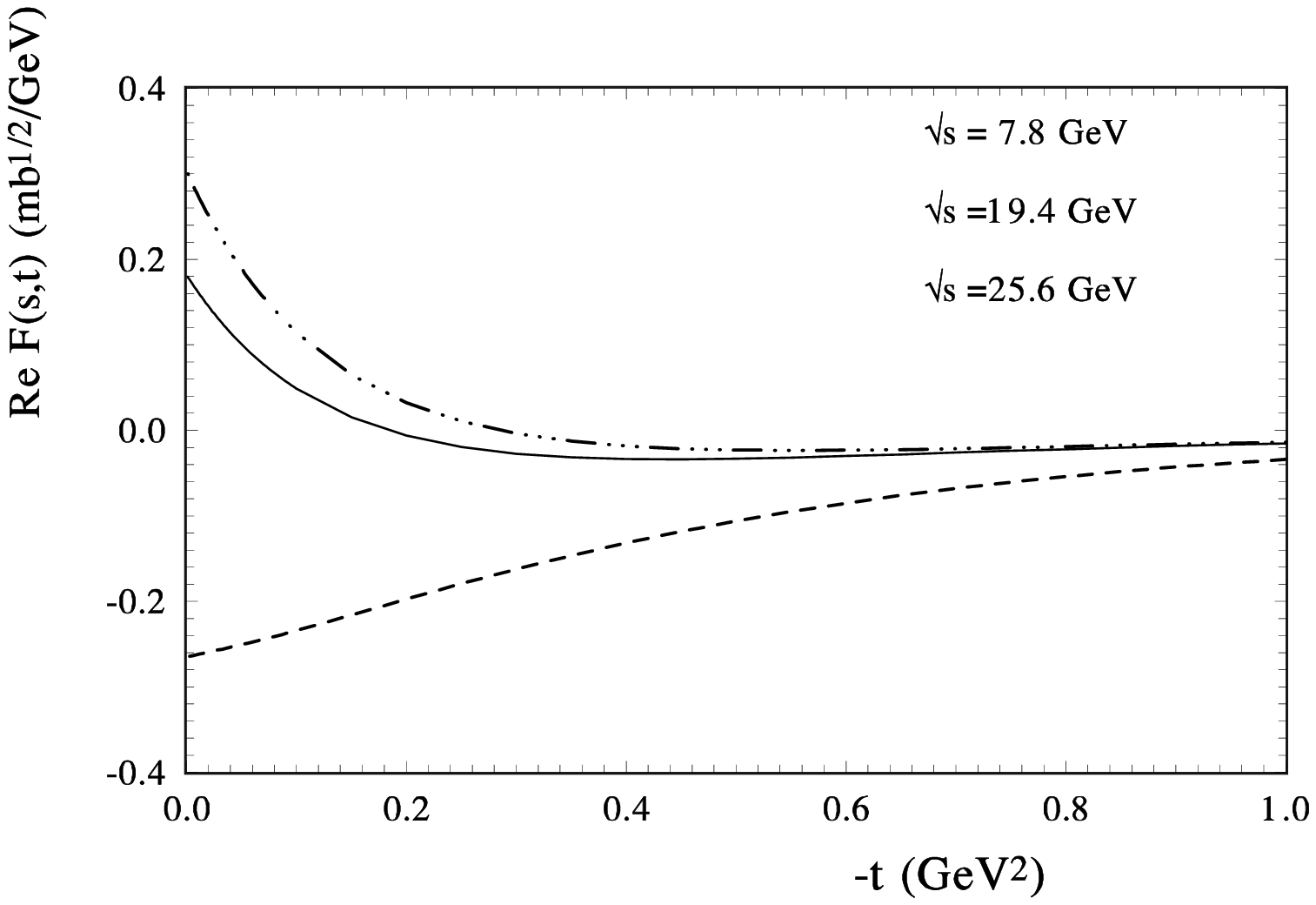} 
\vspace{0.5cm}
\vspace{1.cm}
\caption{ a)  [top] The energy and momentum transfer dependence of the
 imaginary part of the elastic scattering amplitude of  $\pi^{-}p$  and \\
  \hspace{4.cm} b) [bottom] the real part of the elastic scattering amplitude  $\pi^{-}p$
 (the dashed, solid and dotted-dashed  lines
  correspond to the $\sqrt{s} = 25.3, 19.4 \ {\rm and} \ 7.8 \ $ GeV ).
  }
 \end{center}
\label{Fig-11}
\end{figure}

\begin{figure}
\vspace{-1.cm}
\begin{center}
\includegraphics[width=.49\textwidth]{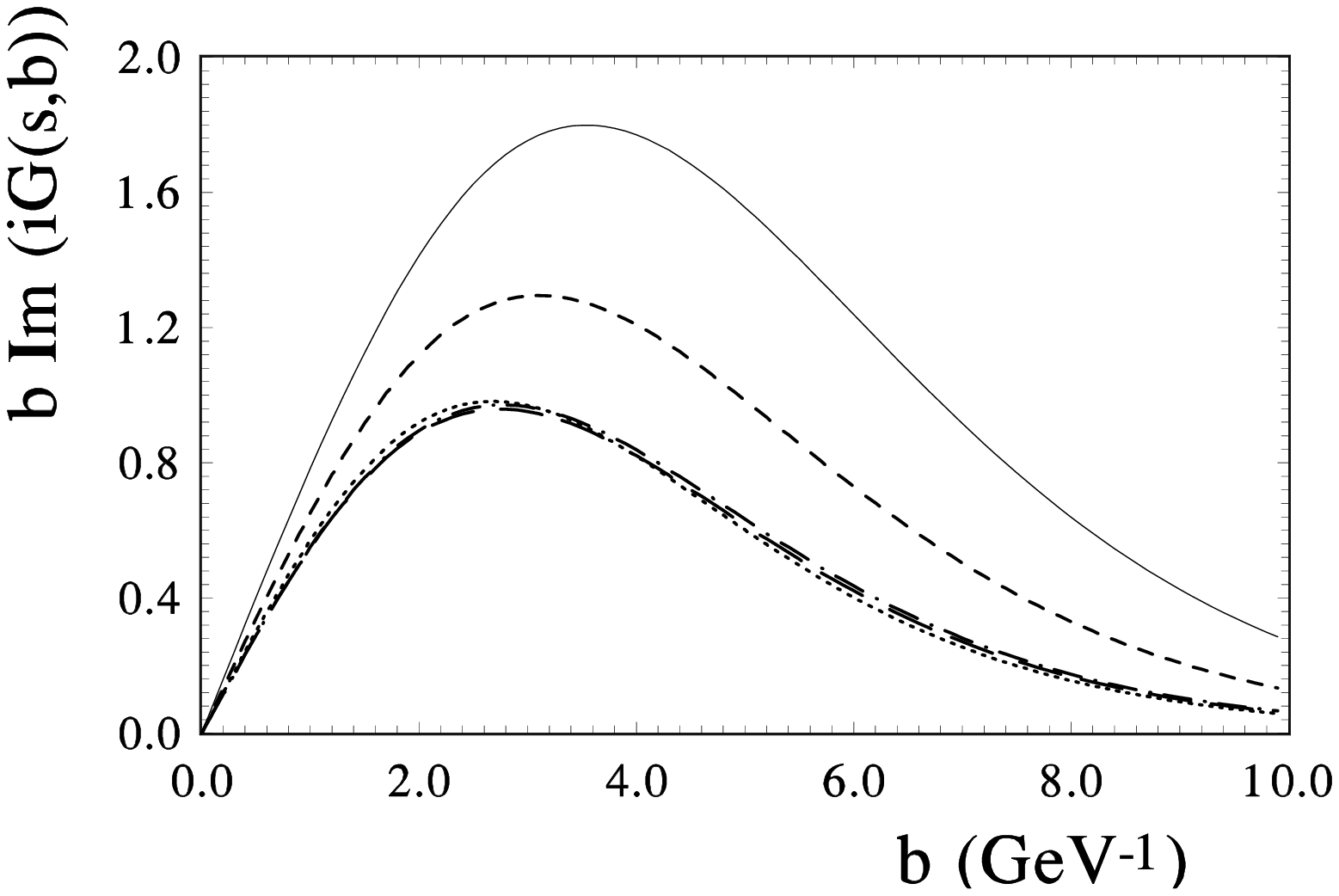} 
\includegraphics[width=.49\textwidth]{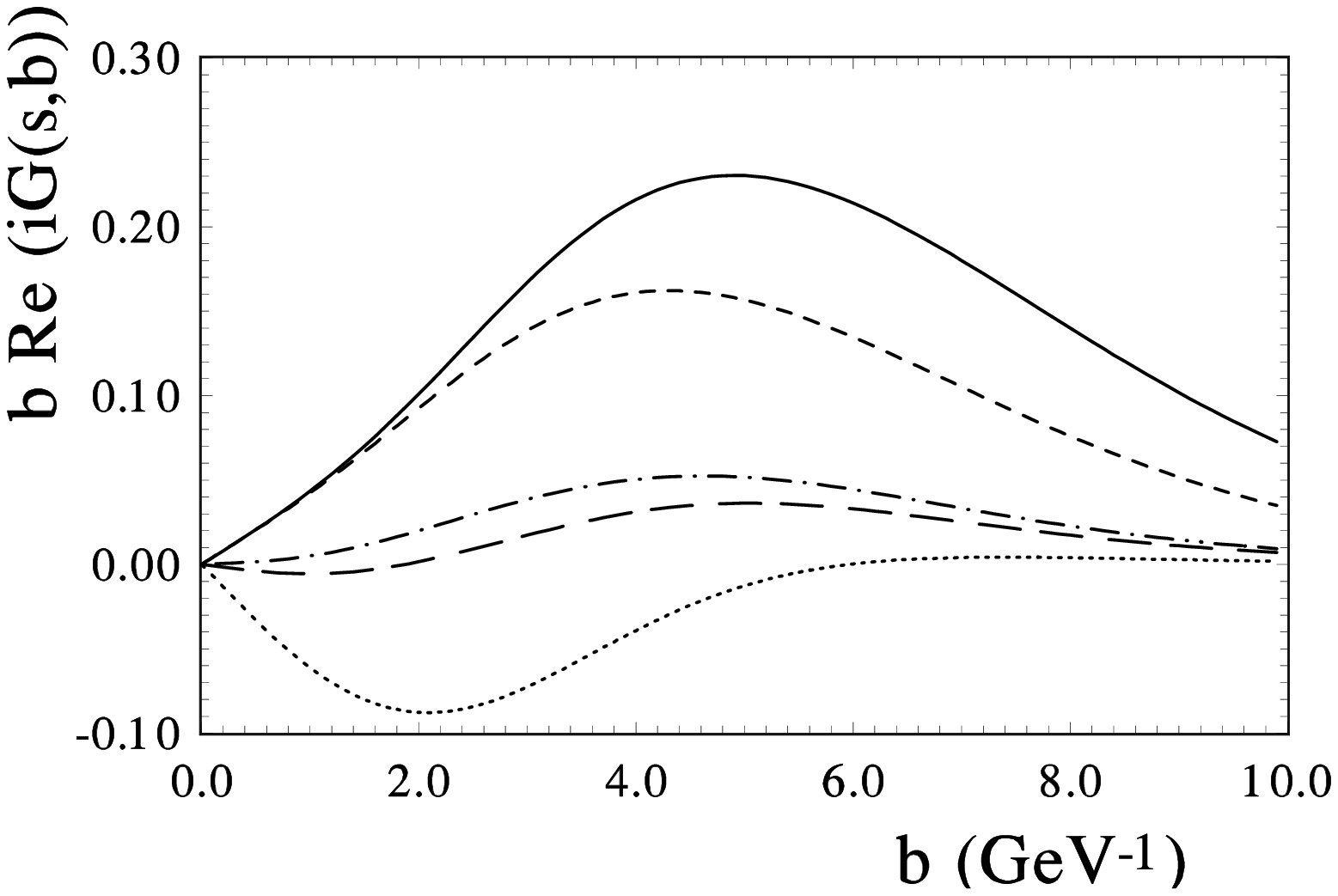} 
\vspace{1.cm}
\caption{ The scattering amplitude  of  $\pi^{-}p$  elastic scattering
  in the impact parameter representation -
  $  i \ b \Gamma(s,b)$
   a) [top] -  the imaginary part and b) [bottom] the real part
  (the points, long-dashed, dotted-dashed, dashed and solid  lines
 correspond to the $\sqrt{s} = 7.8, 19.4, 25.4, 300  \ {\rm and} \ 3000 \ $ GeV ).
  }
 \end{center}
\label{Fig_12}
\end{figure}

\begin{figure}
\begin{center}
\includegraphics[width=.49\textwidth]{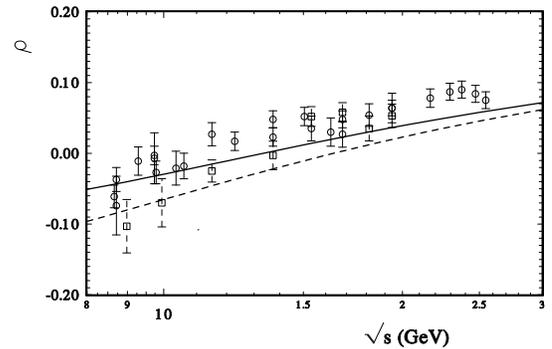} 
\vspace{1.cm}
\caption{ 
 The value $\rho(s,t=0)$ - the ratio of the real to imaginary parts
 of  $\pi^{\pm} p$  elastic scattering amplitude
(curves - our model predictions, circles, squares, triangles up and
 triangles down - \cite{Rub19am,Adam19,Schiz19,Aker19,Cul19,Rub19,Bri19}).
  }
 \end{center}
\label{Fig_13}
\end{figure}

\begin{figure}
\begin{center}
\includegraphics[width=.49\textwidth]{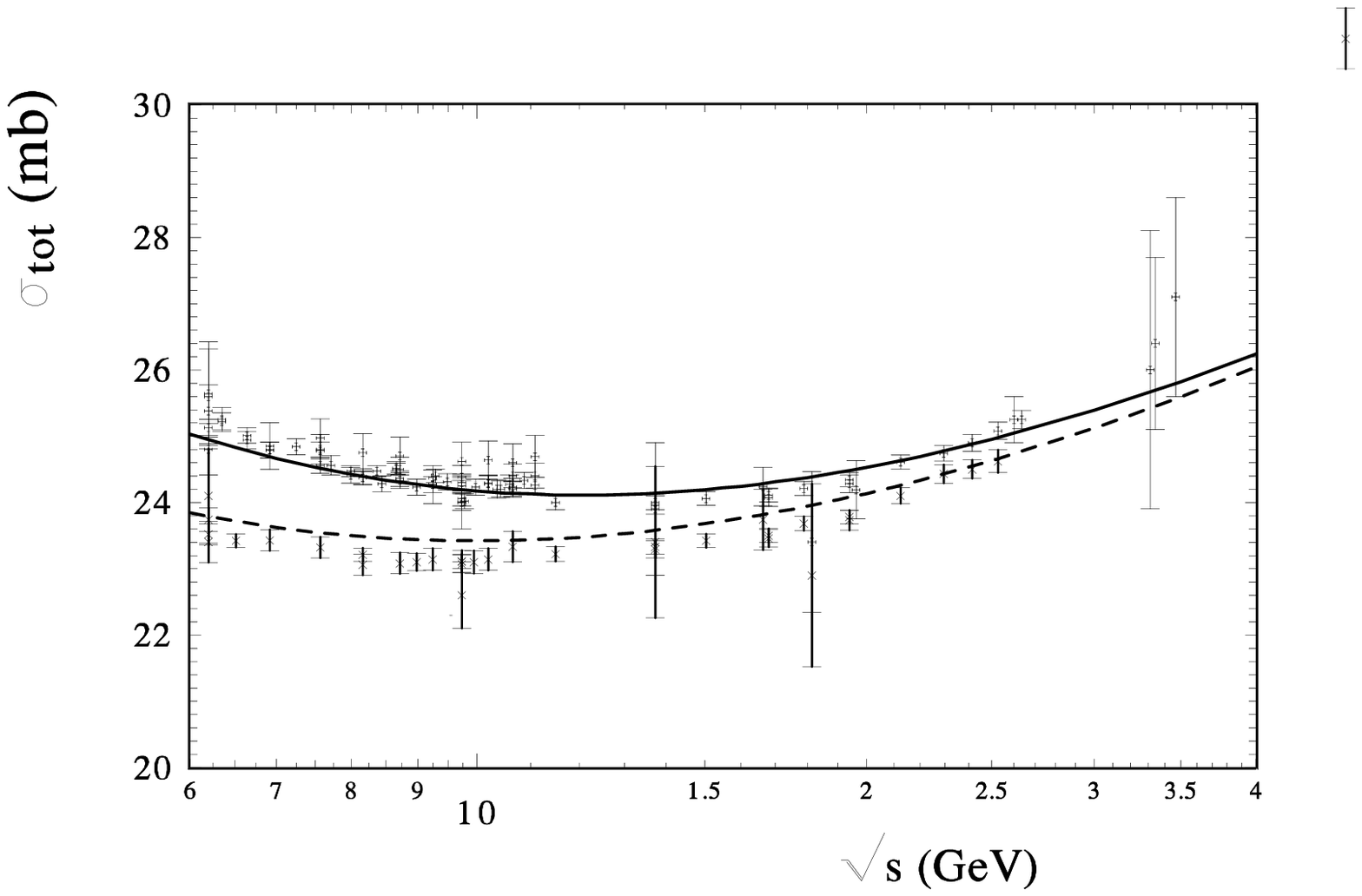} 
\includegraphics[width=.49\textwidth]{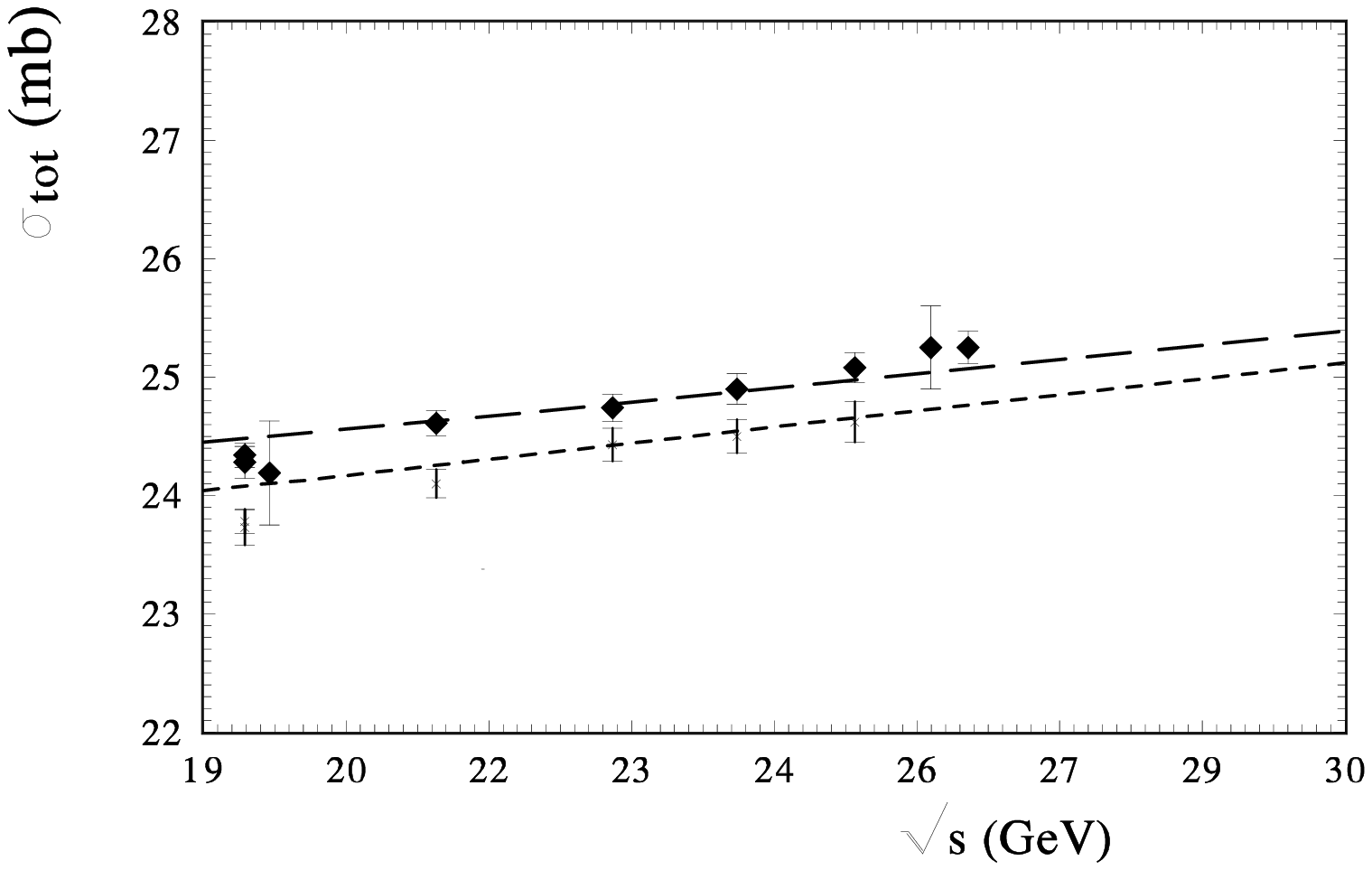} 
\includegraphics[width=.49\textwidth]{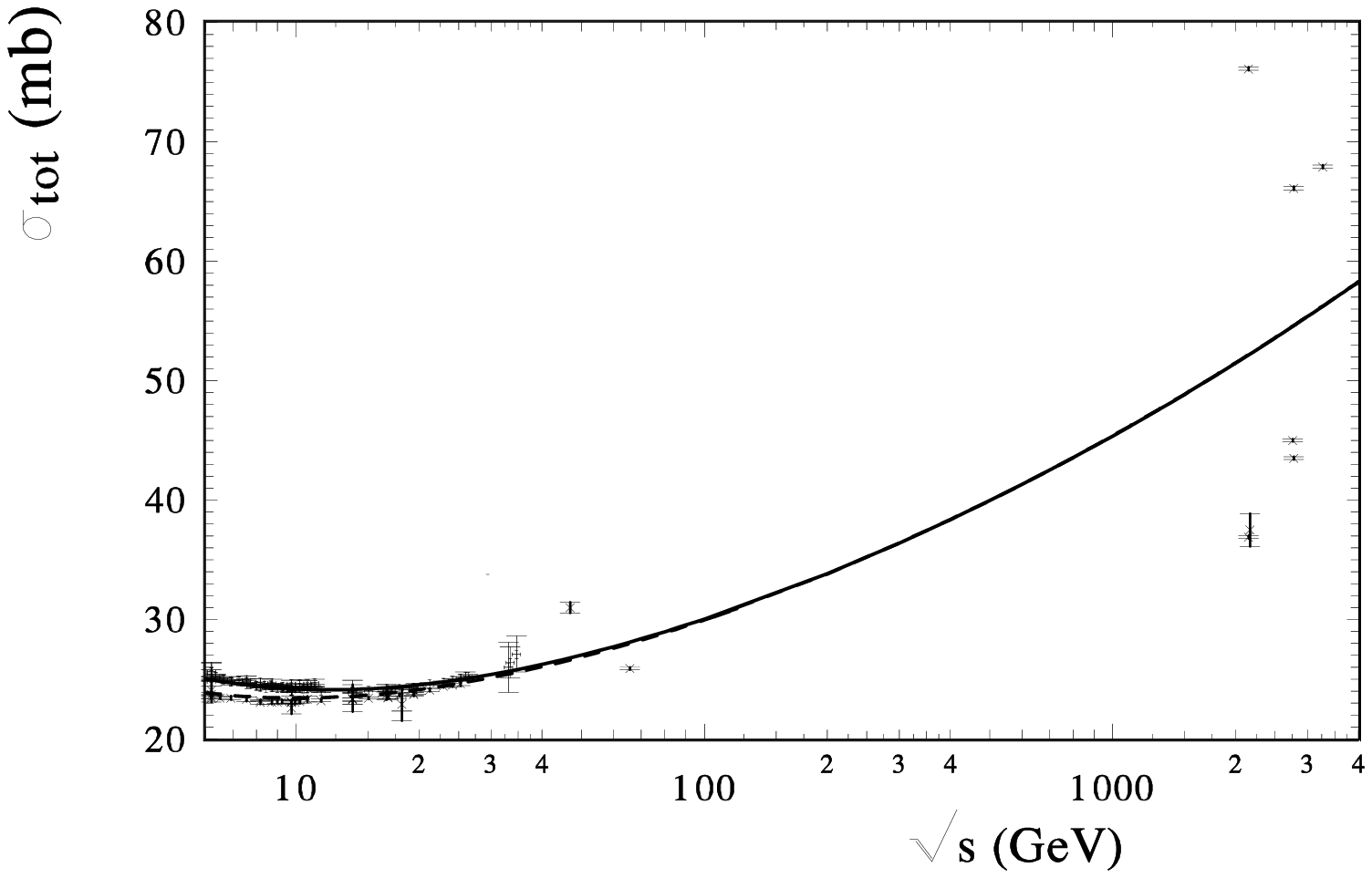} 
\vspace{1.cm}
\caption{
 The total cross sections  of  $\pi^{\pm} p$  elastic scattering
a) at low [top], b) at separate energies [middle] and  c) high  energies [bottom]
(curves - our model predictions, circles, squares, triangles up and
 triangles down - \cite{expdata}). 
  }
 \end{center}
\label{Fig_14}
\end{figure}

 We take into account the experimental data on the  $\pi^{+} p$ and $\pi^{-} p$
    elastic scattering from $\sqrt{s} =  7.807 $ 
    GeV up to the maximum measured
    at $\sqrt{s} = 25.46$ GeV. The total number of the experimental data $N_{exp.}=2009$.
    As in the case of the nucleon scattering, we take into account in
    the fitting procedure the statistical and systematic errors separately.
    Only the statistical errors are included in the standard fitting procedure and
     calculations of
    $\chi^2$. The systematic errors are taken into account as some additional normalization
    of the experimental data of a separate set.
     The whole Coulomb-hadron interference region,
  where the experimental errors are remarkably small,
    was included in our examination of the experimental data
     in the region of momentum transfer $0.00137 \leq |t| \leq 10 $ GeV$^2$.
     After the fitting procedure,
  with the modern version of FUMILY \cite{Sitnik} we
    obtained the total $\sum \chi^2_{i} = 2415$ and
    $(\sum \chi^2)/n_{d.o.f.} = 1.2$  (remember that we used only statistical errors).
    The fitting parameters are obtained as :
    $$h_1 = 0.93 \pm0.01;  \ h_2 = 1.7 \pm 0.02; \ k_1 = 6.7 \pm 0.15; $$
    $$k_2 = 15.7 \pm 0.4; \ \ h_c = 1.4 \pm 0.1 $$
    The model calculations are
    compared with the elastic  $\pi^{-} p$ (Fig.6) and $\pi^{+} p$ (Fig.7)
    at   $\sqrt{s} = 19.4$  GeV. At this energy we have the largest number of experimental
    data in a wide region of momentum transfer. On these figures and others the comparison
    of the experimental data with theoretical calculations is shown with additional normalization
    coefficient equal to unity and with only statistical experimental errors.
    In Fig.8,  such comparison is shown for energy $\sqrt{s} = 25.4 \ $ GeV.
    It is the highest energy at which we have the experimental data
    on  $\pi^{\pm} p$ elastic scattering from the direct $\pi^{\pm} p$ elastic scattering.
    Obviously,  the model gives a good description of the exiting experimental data,
    especially in the small $t$ region where the Coulomb-hadron interference plays an important role.
    The dashed line in Fig.8 shows the model calculations
    at this energy for $\pi^{+} p$ elastic scattering. It can be seen that the  largest difference
   between $\pi^{-} p$ and  $\pi^{+} p$ comes from the Coulomb -hadron interference term
   which has  different signs for these reactions.
   In  Fig.9, the comparison of the model calculations with the experimental data
     is shown at $\sqrt{s} = 9.74$ GeV for $\pi^{-} p$ reactions.
   At last, in Fig.10, the experimental data of $\pi^{+} p$ elastic scattering are compared with the model predictions.
   The data are measured up to $-t= 4  \ $GeV$^2$.
   For this energy the latter value corresponds to  large angles; however, the model describes the data sufficiently well.
   Note that in the figures the comparison of the model results with experimental data
    presented  with only statistical errors and  does not take into account the experimental systematic uncertainty and our additional normalization coefficients.

   The behavior of the imaginary and real parts of the elastic scattering amplitudes
   at different energies  is presented in Fig.11.
  The imaginary parts have a small energy dependence and their momentum transfer
  dependence is practically the same in this energy interval.
   We see  different situations for the real parts of the elastic scattering amplitudes.
  A particularly large difference is shown for  low energies.  It comes from the
  non-asymptotic terms of the scattering amplitude.

   In Fig.12, the elastic scattering amplitude
   $i b \Gamma(s,b)$
    is presented in the impact parameter
   representation at  energies $\sqrt{s} = 7.8, \ 19.4, \ 25.6, \ 300. , \ 3000. $ GeV.
   The imaginary part of the scattering amplitude essentially grows with
   energy and its maximum moves to the biggest value of the impact parameter.
   It reflects the growth of the radius of the hadron interaction.
   Of most  interest is the impact parameter dependence of the real part
   of the scattering amplitude. If at low energy  ($\sqrt{s} = 7.8$)
     its maximum practically
   coincides with the maximum of the imaginary part  (approximately at $2.5$ GeV$^{-1}$),
   then at high energies ($\sqrt{s} = 25.6$ GeV) the positions of the maximum are
   different. The maximum of the imaginary part moves  approximately at
   $3.5$ GeV$^{-1}$), but the maximum of the real part moves at $5.5$ GeV$^{-1}$).
   Such a large difference probably shows  the changes of the hadron potential
   of the interactions at large distances with growing interaction energy.

The experimental data of $\sigma_{tot}(s)$ - the total cross sections and
$\rho(s,t)$ - the ratio of the real to imaginary parts
   of the elastic scattering amplitude at $t=0$ are not included in the fitting procedure.
   These data were extracted from the differential cross sections
   with some simple model representations. Hence, the inclusions of these  data in our
    fitting procedure  will be double account. Let us see what gives the model
    for these values. In Fig. 13, the energy dependence of the $\rho(s,t=0)$ -
    ratio of the real to imaginary parts
 of  $\pi^{\pm} p$  elastic scattering
         is shown.
    It can be see that the model difference between $\rho(s,t=0)_{\pi^{-}p}$
    and  $\rho(s,t=0)_{\pi^{+}p}$  is not large. 
     The model calculations coincide with the experimental data at
   low energy but show  less difference between the reactions
   at high energy. Probably, this is due to the  possible simplification of
   accounting  for the contribution from the second Regions.
    However, in general, the model calculations of $\rho(s,t=0)$ show a good energy dependence
    for both reactions. In Fig.14 a and b, the energy dependence of $\sigma_{tot} $ for these reactions
    is presented at low energies (Fig.14a) and high energies (Fig.14b).
    Obviously, the model  reproduces sufficiently well the energy dependence of  $\sigma_{tot} $
    for both reactions. Note that the last four experimental  data ($\sqrt{s} = 22.5 - 25.4 \ $ GeV)
     for  $\sigma_{tot(\pi^{-}p)} $  usually lie  above the theoretical curves.
     This leads to the opinion of the existence of hard pomeron contributions \cite{CMS}.
     Our HEGS model with only 5 fitting parameters and without taking into account
     the data of $\rho(s,t=0)$ and $\sigma_{tot}(s)$
     in the fitting procedure
     shows that  the hard pomeron contributions  are not  necessary (see Table 1).
     This is consistent with our conclusion  that there is no hard pomeron contribution
     to elastic nucleon-nucleon scattering \cite{NP-hP}.
     In Fig.14b, the model calculations of  $\sigma_{tot(\pi^{\pm}p)}(s) $ are presented with
     experimental data at very large energies. The errors and  distributions of the data
     are very large.
     However, it can be concluded that the model calculations do not contradict
      the recent experimental data.

\begin{table}
\label{Table-1}
\vspace{.5cm}
\caption{The comparison of the model predictions of $\sigma_{tot}(s)$ with the experimental data.
  }
\begin{tabular}{|c|c|c|c|} \hline
  $\sqrt{s}$,GeV &  $\sigma_{tot}^{th}(s) \pm \delta^{th}(s)$, mb&  $\sigma_{tot}^{exp.}(s)$, mb & exper.      \\ \hline
 26.4 &$ 25.0 \pm 0.7$ & $ 25.25 \pm 0.09$ &   \cite{CARROLL-79}  \\
 33.2 &$ 25.5  \pm 0.8$  & $ 26.0  \pm 2.1$  &    \cite{DERSCH-99} \\
 33.4 &$ 25.6  \pm 0.8$  & $ 26.4  \pm 1.3$  &   \cite{DERSCH-99}  \\
 34.7 &$ 25.7  \pm 0.9$  & $ 27.1  \pm 1.5$  &    \cite{DERSCH-99} \\  \hline
\end{tabular}
\end{table}
\vspace{.5cm}

However, in general, the model calculations of $\rho(s,t=0)$ show a good energy dependence
    for both reactions. In Fig.14 a and b, the energy dependence of $\sigma_{tot} $ for these reactions
    is presented at low energies (Fig.14a) and high energies (Fig.14b).
    Obviously, the model  reproduces sufficiently well the energy dependence of  $\sigma_{tot} $
    for both reactions. Note that the last four experimental  data ($\sqrt{s} = 22.5 - 25.4 \ $ GeV)
     for  $\sigma_{tot(\pi^{-}p)} $  usually lie  above the theoretical curves.
     This leads to the opinion of the existence of hard pomeron contributions \cite{CMS}.
     Our HEGS model with only 5 fitting parameters and without taking into account
     the data of $\rho(s,t=0)$ and $\sigma_{tot}(s)$
     in the fitting procedure
     shows that  the hard pomeron contributions  are not  necessary (see Table 1).
     This is consistent with our conclusion  that there is no hard pomeron contribution
     to elastic nucleon-nucleon scattering \cite{NP-hP}.
     In Fig.14b, the model calculations of  $\sigma_{tot(\pi^{\pm}p)}(s) $ are presented with
     experimental data at very large energies. The errors and  distributions of the data
     are very large.
     However, it can be concluded that the model calculations do not contradict
      the recent experimental data.

\section{Conclusion}

  Generalized Parton Distributions reflect the basic properties of the hadron structure
and give  a bridge between  many different reactions.
 We have examined the new form of the momentum transfer dependence of GPDs of hadrons to obtain
       different form factors, including
     Compton form factors, electromagnetic form factors,
     transition form  factor and gravitomagnetic form factors.
Our model of GPDs,
  based on the analysis of  practically all existing experimental data on the
    electromagnetic form factors of the proton and neutron,
 leads to a good description
of the proton and neutron  electromagnetic form factors  simultaneously.
The chosen form of the momentum transfer dependence of GPDs of the pion
(the same as t-dependence of nucleon) allows us to describe the electromagnetic form factor of the pion
and obtain the pion gravitomagnetic form factor.
The obtained parameters of the form factors of the pion and nucleon
satisfy the   quark count.
As a result, the description of different reactions based on
     the same   representation of the hadron structure was obtained.
This especially  concerns  high energy elastic hadron scattering.
The meson High Energy Generalized Structure (mHEGS) model,
taking into account  the electromagnetic and gravitomagnetic form factors
of hadrons,  describes well the
 $\pi^{+}p$ and  $\pi^{-}p$  elastic scattering
in wide energy ($\sqrt{s} > 7$ GeV) and momentum transfer regions
 with a minimum number of fitting parameters, only 5.
The investigation of the nucleon structure  shows that the density of the matter
in hadrons is more concentrated  than the charge density.
     Our calculations show that
      the ratio of the  radii of the  electromagnetic density  to the gravitomagnetic density
     is approximately the same for the nucleon and pion.
The model opens up a new way to determining the true form of the GPDs and
      hadrons structure.

\vspace{0.5cm}
{\bf Acknowledgments}
 {\it The author would like to thank  O.V. Teryaev
   for fruitful   discussions of some questions   considered in the paper.}


\end{document}